# Predicting Near-Future Churners and Win-Backs in the Telecommunications Industry

Clifton Phua, Hong Cao, João Bártolo Gomes, Minh Nhut Nguyen
Data Analytics Department, Institute for Infocomm Research
1 Fusionopolis Way, Connexis,
Singapore 138632

{cwcphua, hcao, bartologjp, mnnguyen}@i2r.a-star.edu.sg

## 1. Introduction

The telecommunications (telco) industry is vibrant and dynamic with a large base of customers. A crucial factor for the survival and good profitability of any telco company is to develop effective strategies to win more customers and to retain the existing customers (e.g. through reducing the percentage of churn). These strategies can help a telco company grow and maintain a large customer base to harvest good profits through economically providing telco services (e.g. voice data transmission and broadband) in a mass scale. However, such strategies cannot be developed without the proper understanding about the reasons why an existing consumer or a Small and Medium Enterprise (SME) chooses to discontinue its telco service plan, i.e. churning. Such timely understanding is particularly important for a company to stay competitive in today's dynamic world with the fast advances of telco technologies, products and services.

Several previous works [1-4] have researched on the prediction of churn through using machine learning and analytical tools. The work in [1] focused on various aspects in developing customer retention strategies. The work in [2] is a recent work and it describes a decision tree classifier when applied to the geographical region of the customer (e.g. rural, sub-urban and urban), plays an important role is the churn prediction. The work in [3] proposed dual-step model building approach, which first clusters different types of customers and then learns the classification models for each cluster separately. It also evaluated cost-sensitive learning technology for churn prediction. The work in [4] differs from the first three works in the following aspects: it not only predicts the churner but also estimates how soon a customer would churn through survival analysis.

The data analytics work presented in this write-up predicts the churners and win-backs in the near future based on recent and large-scale telco data in 2011. Similar to [4], our churn prediction problem becomes more challenging as our technique not only needs to identify the likely churners with good accuracy; our technique also needs to identify these churners within a tight time frame of the subsequent three months. Based on this requirement and considering there could likely be fixed patterns that signal a customer would churn soon in his/her current telco service records, we develop suitable feature computation strategy to extract the relevant features from customer data for predicting the near-future churners and possible win-backs. In building our predictive models, we also perform imbalance correction through sampling techniques with an aim to improve the precision of prediction for the important minority-class. Through comparing different learning algorithms, our method achieves reasonably good prediction performance through cross-validation for the entire customer population. For instance, we achieve good accuracy of 81.8% for SME churn prediction.

## 2. Data Processing and Feature Extraction

We use a telco company's customer data, with both consumers and SMEs, in the full year of 2011. For confidentiality and competitive reasons, the identity of this telco company cannot be revealed. The data was first loaded into two MySQL databases, one for consumers and another for SMEs. Then we created indexes for the key and data fields in order to improve query response time and preprocessing/generation of the training/test sets for the two databases (consumer and SME).

### 2.1 Data Schema

The consumer database (1.3M customers) is significantly larger than the SME one, with close to 15 times more customers. At the heart of both databases are the subscriber tables, which connect to all the other tables with the customer, billing, and service IDs (with the exception of service request tables which connects to customer tables directly through customer ID).

Essentially, each customer (consumer or SME) can have multiple billing accounts, and each billing account can have multiple services. To create flat files for data mining that is reasonably accurate and fast, we opted to have billing ID, instead of customer ID (faster with fewer records) or service ID (more accurate since churn is defined as a single service termination), as the common key to link multiple tables.

Within both consumer and SME databases, the billing tables have the most number of records (15M and 3M), followed by voice usage, service request, subscriber, download and upload usage, and subscriber tables. The class labels for churn and win-back are obtained from the TERMINATION_DATE and COMEBACK_DATE fields in the subscriber table respectively.

### 2.2 Feature Computation Strategy

Given data for the entire 2011 plus class labels from 2011 to January 2012 (churn) and mid-February 2012 (win-back), we divided the data and class labels accordingly:

- To predict churners ('1' as label) and stayers, our strategy involves creating more recent training features from August to October 2011 (3 months). Training class labels are from November 2011 to January 2012 (3 months) to create the model.

- To predict win-back ('1' as label) from churners, our strategy involves creating features for a longer time span (7 months instead of 3) due to the time lag after service termination and because most telco companies run their win-back marketing programs six months after termination date. We determine if the service termination occurs within April and October 2011 (7 months). If so, we create training features for the 3 months preceding the termination month. For example, if a churner terminates a service in June 2011, we extract features for March to May 2011. Training class labels are also from November 2011 to January 2012 (3 months) to learn the churn and comeback prediction models.

- The churn and stay models are subsequently applied on test features created from October to December 2011. The win-back models are subsequently applied on 3-month test features created from service termination within June and December 2011. These models are directly applied to existing customers or churners to predict label-of-interest (churn or win-back) from January to March 2012.

## 2.3 Feature Extraction

This is the most time-consuming part of the data mining process, especially to carefully craft, test, and tune the database queries frequently. We spent the most effort on the service request tables because common sense tells us that number of recent complaints by a subscriber is directly correlated to churn and the database codes in this table changed from May 2011. Surprisingly, the service request features turned out to be the weakest features. On the other hand, the least effort was spent on download usage, upload usage, and billing tables and it turned out that they had the strongest features.

Using shell scripts to collate the final database queries and merge the query output, the original features and class labels are automatically extracted from each table in the two MySQL databases and stitched together using billing ID as common key into flat files. Eventually, there are train and test files to predict consumer churners and stayers, consumer win-back, as well as two types of SME churners and stayers.

In addition to original features, we created some derived features to capture dependencies for the stronger original features. We extracted the differences between original monetary features (for example, CURRENT_BILL_AMT and LAST_BILL_AMT) and between date-time features (for example, ACTIVATION_DATE and CUSTOMER_SINCE), and others.

## 3. Imbalance Correction and Learning

After extracting the features for each customer, we readily form the training and testing datasets based on our earlier mentioned computational strategy. One critical yet common problem we encountered is that our learning set is largely imbalanced and undesirably favors the majority class, which is relatively less important in this study. For instance, the important class of churners always represents a very small percentage, as its size can be hardly comparable to the large population of non-churners. But learning why the existing churners have churned and predicting who will be churners in the near future is the theme of this study. Also in predicting the win-backs, we find that the win-backs in a fixed period of only a few months often represent a relatively insignificant portion as compared with the entire set of churners. As standard learning algorithms often assume balanced datasets and are designed to optimize the overall accuracy, their learned model inevitably favors the majority class. To correct such behavior, we perform imbalance correction through simple sampling strategies and compare the predictive performance of the standard learning techniques implemented in *Weka*.

### 3.1 Imbalance Correction

A large body of imbalanced learning algorithms have been reviewed in a recent survey [5]. In general, one can categorize the mechanisms of imbalance correction into two levels, i.e. algorithm-level and data-level methods. As the former would be tied to specific learning algorithms, we choose the fundamental yet efficient data-level methods for imbalance correction, which can be used together with vast existing learning algorithms. Specifically, we choose both the random under-sampling and repeating oversampling strategies for imbalance correction.

With random under-sampling, we randomly select a subset of the majority-class learning data and use the entire set of minority-class learning samples to form a new balanced learning dataset. Here, balanced learning dataset in our context refers to a two-class learning dataset comprising of equal number of instances from each class. With repeating oversampling, we simply repeat the learning dataset from the minority class multiple times so that the minority class is augmented to a balanced population level of the majority class. Note that many oversampling techniques were proposed in the past [5, 6]. We choose the simple repeating oversampling because our previous imbalanced learning experience [6] finds that it is one of the best techniques to optimize the precision score, which is required learning goal for this challenge.

The under-sampling and oversampling strategies have their respective pros and cons. That is, under-sampling requires discarding the existing instances and this potentially removes informative majority-class instances that are critical to define a good decision boundary between the two classes. On the other hand, though the over-sampling keeps all instances, it would have to increase the size of learning dataset and is therefore less efficient. In this work, we use the under-sampled balanced learning set to perform classification model comparison and best parameter searching. With the best model and parameters found, we retrain our classification model using the oversampled dataset for predicting the most likely set of churners or win-backs in the near future, i.e. for preparation of the required submission files.

### 3.2 The Learning Techniques

There are a large number of standard classification tools we can tap into for making predictions in about six groups, neural networks, Bayesian networks, decision trees, Support vector machines, instanced based and ensemble learning techniques. Especially, previous churn prediction works [1-4] have used the following tools including multi-layer perception neural network, logistic regression, decision tree and survival analysis. In this work, we are particularly interested in the tree-based classifiers for the good interpretability of the learned models. We compare an array of state-of-the-art tree-based classifiers implemented in *Weka*, a popular data mining and machine learning platform. Namely, these tree classifiers include ADTree, Decision Stump, J48, J48graft, TreeLMT, RandomForest, RandomTree, RepTree, SimpleCart, etc. For the promising tree-based classifiers we found, we also use them in conjunction with ensemble learning techniques such as bagging, ClassificationViaRegression and Adaboost. These ensemble-learning frameworks are known to reinforce the learning performance by reducing the performance variations and enhancing the generalization performance.

Table I  Comparison of Different Learning Techniques for Consumer Churn Prediction

|  | ADTree | Decision Stump | RepTree | J48 | NaiveBayes | TreeLMT | Random Forest | Bagging +Decision Stump | Bagging +Simple Cart | SimpleCart | Clas.Via .Regression |
|---|---|---|---|---|---|---|---|---|---|---|---|
| **Prec_1.** | 75.8 | **86.6** | 75.8 | 69.2 | 74.0 | 77.7 | 75.7 | 86.6 | 74.5 | 77.4 | 78.6 |
| **Prec_0** | 70.8 | 59.7 | 71.4 | 70.0 | 69.1 | 71.9 | **69.7** | 59.7 | 72.7 | 72.2 | 71.0 |
| **Accu.** | 73.0 | 65.3 | 73.4 | 69.6 | 71.3 | 74.5 | **72.3** | 65.4 | 73.6 | 74.5 | 74.2 |

**Prec_1**: Precision for churner prediction; **Prec_0**: Precision for non-churner prediction
**Accu.**: Accuracy rate for classification of churners and non-churners

Table II  Comparison of Different Learning Techniques for Consumer Win-Back Prediction

|  | ADTree | Decision Stump | RepTree | J48 | NaiveBayes | TreeLMT | Random Forest | Bagging +BF Tree | Bagging +LadTree | SimpleCart | Clas.Via .Regression |
|---|---|---|---|---|---|---|---|---|---|---|---|
| **Prec_1.** | 67.7 | 59.8 | 66.0 | 63.2 | 64.6 | 68.1 | 67.3 | 65.3 | 67.1 | 67.8 | **68.8** |
| **Prec_0** | 66.0 | 64.7 | 65.5 | 62.7 | 63.9 | 67.7 | 63.2 | 66.8 | 67.2 | **69.6** | 67.7 |
| **Accu.** | 66.8 | 60.7 | 65.8 | 63.0 | 64.2 | 67.9 | 65.0 | 66.0 | 67.2 | 67.2 | **68.2** |

**Prec_1**: Precision for win-back prediction; **Prec_0**: Precision for churner prediction
**Accu.**: Accuracy rate for classification of win-backs and churners

Table III  Comparison of Different Learning Techniques for SME Churn Prediction for Voice Services Only

|  | ADTree | Decision Stump | FT | J48 | J48graft | LADTree | NBTree | RandomForest | RandomTree | RepTree | SimpleCart | Clas.Via.Regression |
|---|---|---|---|---|---|---|---|---|---|---|---|---|
| **Prec_1.** | 80.3 | **93.4** | 80.3 | 78.1 | 78.3 | 79.5 | 74.3 | 83.9 | 76.3 | 78.7 | 77.0 | 81.8 |
| **Prec_0** | 71.0 | 63.9 | 71 | 76.5 | 76.6 | 72.4 | 73.0 | **79.9** | 73.7 | 73.1 | 75.5 | 75.3 |
| **Accu.** | 74.8 | 71.1 | 74.8 | 77.3 | 77.4 | 75.5 | 73.6 | **81.8** | 74.8 | 75.6 | 76.2 | 77.1 |

**Prec_1**: Precision for churner prediction; **Prec_0**: Precision for non-churner prediction
**Accu.**: Accuracy rate for classification of churners and non-churners

Table IV  Comparison of Different Learning Techniques for SME Churn Prediction for Voice & Broadband Services

|  | ADTree | Decision Stump | FT | J48 | J48graft | LADTree | NBTree | RandomForest | Bagging Simple Cart | Bagging +RepTree | Simple Cart | Clas.Via.Regression |
|---|---|---|---|---|---|---|---|---|---|---|---|---|
| **Prec_1.** | 73.5 | 74 | 67.5 | 70.6 | 71.1 | **77.4** | 68.8 | 75.3 | 74.2 | 71.9 | 72.4 | 70.2 |
| **Prec_0** | 67.2 | 62.7 | 67.9 | 70.5 | 70.8 | 65.6 | 68.8 | 71.0 | 72.7 | 70.6 | 70.1 | **71.4** |
| **Accu.** | 69.8 | 66.6 | 67.7 | 70.5 | 70.9 | 69.8 | 68.8 | 73.0 | **73.4** | 71.2 | 71.2 | 70.8 |

**Prec_1**: Precision for churner prediction; **Prec_0**: Precision for non-churner prediction
**Accu.**: Accuracy rate for classification of churners and non-churners

## 4. Problem definition

Here is a list of prediction problems we have considered.
1. Predict potential **consumer** churners.
2. Predict most loyal **consumer**s – those who are least likely to churn.
3. Predict **consumer** win-backs - ex-consumers who are likely to come back as customers.
4. Predict potential **SME** churners for voice services only.
5. Predict most loyal **SME** customers for voice services only.
6. Predict potential **SME** churners subscribing both voice *and broadband* services.
7. Predict most loyal **SME** customers subscribing both voice *and broadband* services.

## 5. Experimental Results

We use the under-sampled data obtained from Month 8 to Month 10 as our training data. We perform 10-fold experiments on a list of classifiers to find the best classier which achieves the highest precision result. The list of classifiers includes: LADTree, Decision Stump, RepTree, J48, NaiveBayes, TreeLMT, RandomForest, Bagging+BF Tree, Bagging+LADTree, SimpleCart, ClassificationViaRegression. Based on the comparison results in Table I to IV, we select and apply the best learning algorithm for each problem, as shown in Table V.

Besides the comparison results, we have also tabulated the top-15 features in Table VI that best predict the near-future churners. It shows the utilization and payment patterns in the most recent months that are highly indicative of whether a consumer would churn soon. As an example, we have also illustrated an Alternate Decision Tree (ADTree) model in Fig. 1 that we have learned from the consumer churn prediction with several other influential features used.

Table V  Best Algorithm and Its Result for Each Problem

| Problem (see Section 4) | Classification Algorithm | Precision (%) |
|---|---|---|
| 1 | SimpleCart | 77.4 |
| 2 | SimpleCart | 72.2 |
| 3 | LADTree | 67.7 |
| 4 | Decision Stump | 93.4 |
| 5 | RandomForest | 79.9 |
| 6 | LADTree | 77.4 |
| 7 | ClassificationViaRegression | 71.4 |

Table VI  Ranking of Top-15 Features that Best Predicts Consumer Churn

| Feature | Description |
|---|---|
| DL1110 | Download volume in Oct (in MB) |
| UL1110 | Upload volume in Oct (in MB) |
| 3M_DL_avg | Average download volume for 3 months (Aug, Sep, Oct) |
| 3M_UL_avg | Average upload volume for 3 months (Aug, Sep, Oct) |
| DL1109 | Download volume in Sep (in MB) |
| UL1109 | Upload volume in Sep (in MB) |
| AMT_2PAY_avg | Average amount to pay (Aug, Sep, Oct) |
| OUTSTANDING_avg | Average outstanding amount (Aug, Sep, Oct) |
| PAYMENT_avg | Average payment made (Aug, Sep, Oct) |
| DIFF_AMT_2PAY_PRICE_START | Difference between AMT_2PAY_avg and Price_Start |
| DIFF_CURRENT_LAST_BILL_AMT_avg | Difference between CURRENT_BILL_AMT_avg and LAST_BILL_AMT_avg |
| LAST_BILL_AMT_avg | Average last month bill amount (Aug, Sep, Oct) |
| DL1108 | Download volume in Aug (in MB) |
| UL1108 | Upload volume in Aug (in MB) |
| DL1107 | Download volume in Jul (in MB) |

```
Problem 1 (Churner Prediction) – ADTree Model
: 0
| (1)UL1110 < 0.5: 0.941
| | (3)OUTSTANDING_avg < 442.5: -0.318
| | (3)OUTSTANDING_avg >= 442.5: 0.512
| | | (10)CREDIT_ADJ_avg < -9.5: 0.751
| | | (10)CREDIT_ADJ_avg >= -9.5: -0.281
| (1)UL1110 >= 0.5: -0.196
| | (2)HSBB_Area < 0.5: -0.183
| | | (6)T_Location = AJP: 0.697
| | | (6)T_Location != AJP: -0.036
| | | | (9)T_Location = TLS: 0.585
| | | | (9)T_Location != TLS: -0.006
| | (2)HSBB_Area >= 0.5: 0.691
| (4)ACTIVATION_DATE_TENURE < 15.5: -0.624
| (4)ACTIVATION_DATE_TENURE >= 15.5: 0.042
| (5)Contract_Period < 21: 0.063
| (5)Contract_Period >= 21: -0.265
| | (7)ACTIVATION_DATE_TENURE < 29.5: -0.388
| | (7)ACTIVATION_DATE_TENURE >= 29.5: 0.175
| (8)PAYMENT_avg < -38.5: -0.039
| (8)PAYMENT_avg >= -38.5: 0.284
```

Fig. 1  Alternate Decision Tree (ADTree) Model for Consumer Churn Prediction

## 6. Discussion

In this work, we presented the strategies and techniques that we have developed for predicting the near-future churners and win-backs for a telecom company. On a large-scale and real-world database containing customer profiles and some transaction data from a telecom company, we first analyzed the data schema, developed feature computation strategies and then extracted a large set of relevant features that can be associated with the customer churning and returning behaviors. Our features include both the original driver factors as well as some derived features. We evaluated our features on the imbalance corrected dataset, i.e. under-sampled dataset and compare a large number of existing machine learning tools, especially decision tree-based classifiers, for predicting the churners and win-backs. In general, we find RandomForest and SimpleCart learning algorithms generally perform well and tend to provide us with highly competitive prediction performance. Among the top-15 driver factors that signal the churn behavior, we find that the service utilization, e.g. last two months' download and upload volume, last three months' average upload and download, and the payment related factors are the most indicative features for predicting if churn will happen soon. Such features can collectively tell discrepancies between the service plans, payments and the dynamically changing utilization needs of the customers. Our proposed features and their computational strategy exhibit reasonable precision performance to predict churn behavior in near future.